\documentclass[prl,twocolumn,superscriptaddress,showpacs,floatfix]{revtex4}

\usepackage[final]{graphicx}
\usepackage{amsfonts,amssymb,amsmath}

\newcommand{\mrm}[1]{\ensuremath{\mathrm{#1}}}
\newcommand{\hc}[1]{\ensuremath{#1^\dagger}}

\newcommand{\bs}[1]{\ensuremath{\boldsymbol{#1}}}
\newcommand{\cd}{\cdot}
\newcommand{\al}[1]{\begin{align} #1 \end{align}}
\newcommand{\bT}[1]{\ensuremath{\left\{ #1 \right\}}}
\newcommand{\bP}[1]{\ensuremath{\left( #1 \right)}}
\newcommand{\bS}[1]{\ensuremath{\left[ #1 \right]}}
\newcommand{\e}[1]{\mathrm{e}^{#1}}
\newcommand{\abs}[1]{\lvert #1 \rvert}
\newcommand{\pd}[1]{\partial_{#1}}
\newcommand{\sig}[1]{\sigma_{#1}}
\newcommand{\om}[1]{\omega_{#1}}
\newcommand{\hb}{\hbar}
\newcommand{\gamot}{\gamma_{12}}
\newcommand{\GG}{\mathcal{G}}

\begin{document}

\title{Non-Markovian model of photon-assisted dephasing by electron-phonon interactions in a coupled quantum-dot-cavity system}

\author{P. Kaer}\email{per.kaer@gmail.com}
\affiliation{DTU Fotonik, Department of Photonics Engineering, Technical University of Denmark, Building 345, 2800 Kgs. Lyngby, Denmark}
\author{T. R. Nielsen}
\affiliation{DTU Fotonik, Department of Photonics Engineering, Technical University of Denmark, Building 345, 2800 Kgs. Lyngby, Denmark}
\author{P. Lodahl}
\affiliation{DTU Fotonik, Department of Photonics Engineering, Technical University of Denmark, Building 345, 2800 Kgs. Lyngby, Denmark}
\author{A.-P. Jauho}
\affiliation{DTU Nanotech, Department of Micro- and Nanotechnology Engineering, Technical University of Denmark, Building 344, 2800 Kgs. Lyngby, Denmark}
\affiliation{Department of Applied Physics, Helsinki University of Technology, P. O. Box 1100, FI-02015 HUT, Finland}
\author{J. M{\o}rk}
\affiliation{DTU Fotonik, Department of Photonics Engineering, Technical University of Denmark, Building 345, 2800 Kgs. Lyngby, Denmark}

\date{\today}

\begin{abstract}
We investigate the influence of electron-phonon interactions on the
dynamical properties of a quantum-dot-cavity QED system. We show
that non-Markovian effects in the phonon reservoir lead to
strong changes in the dynamics, arising from
photon-assisted dephasing processes, not present in Markovian
treatments. A pronounced consequence is the emergence of a phonon
induced spectral asymmetry when detuning the cavity from the
quantum-dot resonance. The asymmetry can only be explained when
considering the polaritonic quasi-particle nature of the
quantum-dot-cavity system. Furthermore, a temperature induced
reduction of the light-matter coupling strength is found
to be relevant in interpreting experimental data, especially in
the strong coupling regime.
\end{abstract}

\pacs{78.67.Hc, 03.65.Yz, 42.50.Pq}

\maketitle


The emergent field of quantum information technology
\cite{Knill.Nature.2001} has spurred major research activities on
controlling the fundamental interaction between a semiconductor
quantum-dot (QD) and a cavity. Solid-state cavity QED (cQED) systems
are inherently coupled to the environment, since the emitter is
embedded in a solid. This is in contrast to atomic cQED where the
atom can be effectively isolated and only few discrete energy levels
are sufficient in the description. Remarkably dephasing from
solid-state environments cannot simply be seen as a nuisance, but
can in fact lead to enhanced coupling of QDs to a detuned cavity
mode of importance for efficient single-photon sources
\cite{LivelyDebate,FirstStrongCoup,PureDephPhen} and nanolasers
\cite{Strauf.PRL.2006}. Modeling the continuum of reservoir modes of
solid-state systems constitutes a considerable challenge. The
coupling of the QD-cavity system to its solid-state environment has
almost exclusively been described using Markovian theories
\cite{Carmichael.PRA.1989,PureDephPhen}, neglecting memory effects
of the reservoirs. While the Markovian approximation is well
justified for some reservoirs, this is not in general true for the
reservoir consisting of quantized lattice vibrations. Such phonon
reservoirs dephase the QD-cavity system, whereby the entanglement
between light and matter in general is destroyed. Notably the first
experimental demonstrations of the strong coupling regime in
solid-state cQED \cite{FirstStrongCoup} revealed features in the
emission spectra for large QD-cavity detuning that could not be
explained by standard Markovian theory \cite{Carmichael.PRA.1989}.
Since then there has been a lively debate
\cite{LivelyDebate,Winger.arXiv.2009,PureDephPhen} on the origin of
the deviations. We demonstrate that non-Markovian phonon processes
play an important role for solid-state cQED.

Here, using
a simple physical model we show that
photon-assisted dephasing processes are
of great importance in describing the effect of phonons in a cQED
setting. The underlying physical picture is that the polariton
quasi-particle, formed by dressing the QD with the cavity photon, is
dephased by phonon processes. We focus on the regime of relatively
small QD-cavity detunings and pulsed excitation conditions where
dephasing processes mediated by longitudinal acoustic (LA) phonons
are expected to be important, and investigate the consequences on
the dynamical properties of the cQED system. Pulsed excitation is
required for on-demand photon sources emitting indistinguishable
single photons and entangled-photon pairs \cite{Knill.Nature.2001},
making it an important regime to investigate. Our theory  takes into
account memory effects of the phonon bath, which are neglected in
the usual Markovian Lindblad theory of dephasing processes.
The non-Markovian formulation is shown to be vital in interpreting
recent experiments \cite{Winger.arXiv.2009}.

A similar non-Markovian theory has recently been used to describe
the field dependent dephasing dynamics of classically driven
two-level systems without a cavity
\cite{Classical}.
By varying the strength of the applied classical field one can
approach a regime where the QD dynamics takes place on a time scale
that is near or even below the correlation time of the phonon
reservoir, where the usual Markovian Lindblad theory of decay
breaks down. In cQED the coupling is mediated by a single
photon with a coupling strength $g$. State-of-the-art samples
\cite{Winger.arXiv.2009} have coupling strengths up to $\hb g =
150~\mu$eV, translating to a characteristic time scale of about 14
ps, which is considerably longer than the phonon reservoir
correlation time of typically 3-5 ps. Importantly, even for these
realistic parameters non-Markovian effects are found to play an
important role, giving rise to non-trivial detuning dependent
dynamics and coupling strength. The QD-LA phonon interaction and its
effects on spectra and dynamics are well understood in the
semiclassical regime \cite{semiclasQDLA}, whereas effects due to
quantized light fields have not received much attention. Initial
work on the influence of phonon dephasing on a cQED system has been
reported \cite{initialWork}.
Common to these works is that little physical insight into the
non-Markovian nature of the dephasing processes has been given,
which is a central theme of this manuscript. We note that  the
physical processes identified here are also expected to be relevant
to other cQED systems coupled to vibrational reservoirs, e.g.
quantum wells, organic molecules, nitrogen vacancy centers in
diamond, and colloidal QDs \cite{OtherTLS}, implying that the ideas
presented here may have wide applications.

We describe the effect of LA phonons on the
QD-cavity system using the Jaynes-Cummings model with the addition
of the electron-LA phonon interaction
\cite{QEDLAphonon,Milde.PRB.2008}. The QD has an excited and a
ground state of energies $\hb\om{\mrm e}$ and $\hb\om{\mrm g}$,
respectively, while the cavity photon has energy
$\hb\om{\mrm{cav}}$. The QD-cavity system space is spanned by the
two-level basis $\bT{ | 1 \rangle= | \mrm e,n=0 \rangle, | 2 \rangle
= | \mrm g,n= 1 \rangle}$, where $n$ is the cavity occupancy, thus
$| 1 \rangle$ describes the excited QD and $| 2 \rangle$ describes
the excited cavity. The Hamiltonian has three terms,
$H=H_\mrm{s}+H_\mrm{i}+H_\mrm{ph}$, described below. The Hamiltonian
of the QD-cavity system is $H_\mrm{s}=\hb \Delta\sig{11}+\hb
g(\sig{12}+\sig{21})$, where $\Delta
=\om{\mrm{e}}-\om{\mrm{g}}-\om{\mrm{cav}}$ is the QD-cavity detuning
and $\sig{nm}=| n\rangle\langle  m|$. The electron-phonon
interaction is $H_\mrm{i}=\sig{11}\sum_{\bs k}M^{\bs k}(\hc b_{-\bs
k}+b_{\bs k})$
, where $M^{\bs k}=M^{\bs k}_\mrm{ee}-M^{\bs k}_\mrm{gg}$ is the
effective phonon interaction matrix element \cite{phononMEdef} and
$\hc b_{\bs k}$ creates a phonon in mode $\bs k$. The free phonon
Hamiltonian is $H_\mrm{ph}=\sum_{\bs k}\hb \om{\bs k} \hc b_{\bs k}
b_{\bs k}$, where $\om{\bs k} = c_\mrm s k$ is the phonon dispersion
and $c_\mrm s$ is the speed of sound.

We apply the timeconvolution-less approach
\cite{OpenSys.Breuer.Petruccione.2002} for describing the reduced
dynamics of the QD-cavity system.
The density matrix of the QD-cavity system, $\rho(t)$, is considered
to second order in $H_\mrm i$,
and assuming that the phonon bath remains in a thermal state.
The equations for $s_{pq}(t)=\mrm{Tr}\{\rho(t)\sig{pq}\}$
(representing the excited QD state population, $s_{11}(t)$, the
number of photons in the cavity, $s_{22}(t)$, and the
photon-assisted polarization, $s_{12}(t)$) are
\begin{subequations}\label{eq:DM_eoms}
\al{
&\pd t s_{11}(t) = -ig\bS{s_{12}(t)-s^*_{12}(t)}-\Gamma s_{11}(t),\\
&\pd t s_{22}(t) = ig\bS{s_{12}(t)-s^*_{12}(t)}-\kappa s_{22}(t),\\\notag
&\pd t s_{12}(t) = i\Delta s_{12}(t)-ig\bS{s_{11}(t)-s_{22}(t)}\\\label{eq:s12_eom}
&\quad\quad\quad\quad\quad\quad-1/2(\Gamma+\kappa)s_{12}(t)+\pd t s_{12}(t)\vert_\mrm{ph}.
}
\end{subequations}
Without the phonon induced terms represented by $\pd t
s_{12}(t)\vert_\mrm{ph}$ these equations are the standard lossy
Jaynes-Cummings model. The losses have been introduced through the
Lindblad formalism \cite{OpenSys.Breuer.Petruccione.2002}. These
include decay, described by a rate $\Gamma$, of the excited QD state
to modes other than the cavity and non-radiative channels, and the
finite linewidth of the cavity, $\kappa  = \om{\mrm{cav}}/Q$, where
$Q$ is the usual quality factor of the cavity. We take $\Gamma  =
1~\mrm{ns}^{-1}$ in all simulations to be presented. The phonon
induced terms in Eq. (\ref{eq:s12_eom}) are
\al{\notag &\pd t
s_{12}(t)\vert_\mrm{ph} =
-\bS{\gamma_{12}(t)-i\Delta_\mrm{pol}}s_{12}(t)\\\label{eq:s12_ph}
&\quad\quad\quad\quad\quad\quad\quad\quad+i\mathcal{G}^<(t)s_{22}(t)-i\mathcal{G}^>(t)s_{11}(t).
}
This term introduces two novel effects compared to standard
Markovian cQED models \cite{PureDephPhen}. Firstly, $\gamma_{12}(t)$
enters as a time-dependent pure dephasing rate. Secondly, the
functions $\mathcal{G}^\gtrless(t)$ renormalize the bare coupling
strength $g$: the effective value of $g$ is changed by the real part
of $\GG^\gtrless(t)$, and an additional decay of the polarization is
induced by the imaginary part of $\GG^\gtrless(t)$.
The long-time polaron shift $\Delta_\mrm{pol} =
\mrm{Im}\bT{ \gamma_{12}(\infty) }$ has been subtracted from
$\gamma_{12}(t)$ \cite{OpenSys.Breuer.Petruccione.2002}.
Explicitly, \al{\label{eq:Ggl_def}
\mathcal{G}^\gtrless(t)&=i\hb^{-2}\int_0^t dt'
U^*_{11}(t')U_{21}(t')D^\gtrless(t'),\\\label{eq:gam12_def}
\gamma_{12}(t) &= \hb^{-2}\int_0^t dt'[\abs{U_{11}(t')}^2D^<(t')
-\abs{U_{21}(t')}^2D^>(t')]. } The phonon bath correlation functions
are \al{\label{eq:Dgtrless_def} D^\gtrless(t)=\sum_{\bs k}
\abs{M^{\bs k}}^2 \bS{n_{\bs k}\e{\pm i \om{\bs k}t}+\bP{n_{\bs
k}+1}\e{\mp i \om{\bs k}t}}, } with $n_{\bs k} = 1/(\exp(\hb \om{\bs
k}/k_B T)-1)$. The operator $U(t)=\exp (-iH_\mrm{s} t/\hb)$ is the
time evolution operator for the QD-cavity system and is the
essential ingredient 
as it introduces the
photon-dressed QD into the phonon scattering terms. The 
Markovian Lindblad formalism is obtained neglecting all
memory effects associated with the phonon interaction,
i.e. $D^\gtrless(t) \propto \delta(t)$
\cite{OpenSys.Breuer.Petruccione.2002}. In this limit
$\GG^\gtrless(t)=0$ and $\gamma_{12}(t)=\mrm{constant}$, with no
dependence on photon properties.

\begin{figure}[t]
  \includegraphics[width=0.4\textwidth]{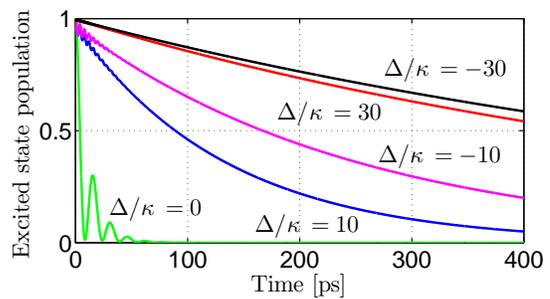}
\caption{(color online). Evolution of the excited state population $s_{11}(t)$ for $\hb g = 140~\mu$eV, $\hb \kappa = 100~\mu \mrm{eV}$, $Q \approx 10^4$, and $T = 4$ K.}
  \label{fig:Purcell_decay_curves}
\end{figure}
We have numerically solved Eqs. (\ref{eq:DM_eoms}) with the initial
condition of a single excitation on the QD, $s_{11}(0)=1$,  and all
other elements set to zero,
modeling an experiment where a QD is excited
in its discrete states with a short optical pulse \cite{ExciteCon}. The parameters are
chosen similar to recent experiments \cite{Winger.arXiv.2009}. The
computed excited state populations, $s_{11}(t)$, are shown in Fig.
\ref{fig:Purcell_decay_curves} for various detuning values.
For $\Delta = 0$ we observe an expected strong
enhancement of the decay rate and associated Rabi oscillations,
indicating the strong coupling regime. For non-zero detuning we
observe small Rabi oscillations in the start of the decay curve,
again indicative of the strong coupling regime. Interestingly, our
theory predicts a shorter lifetime when the cavity is tuned below
the QD resonance than when it is tuned above, in stark contrast with
standard Markovian theory \cite{Andreani.PRB.1999}. The asymmetry is
particularly strong for the $\Delta/\kappa=\pm 10$ cases and has
recently been observed in experiments \cite{Winger.arXiv.2009}. We
attribute this pronounced reduction in lifetime to a phonon-assisted
Purcell effect, where the QD may couple via phonon emission to the
cavity when the cavity is spectrally below it. In the opposite case,
where the cavity is spectrally above the QD resonance, phonon absorption is
needed for the QD to become resonant with the cavity. This process is
suppressed at low temperatures and the phonon mediated coupling
between QD and cavity is lost. The asymmetry becomes less pronounced
when the detuning is larger than the energy the electron can lose
through phonon emission, as seen in the $\Delta/\kappa = \pm 30$
cases. The importance of using the photon-assisted electron-phonon
interaction,
(i.e., the operator $U(t)$ in Eqs. (\ref{eq:Ggl_def}) and
(\ref{eq:gam12_def})), should be emphasized. Setting $g=0$ only in
$U(t)$ we obtain results in quantitative agreement with the $\Delta
< 0$ results in Fig. (\ref{fig:Purcell_decay_curves}), where the
phonons do not significantly influence the dynamics, for all values
and signs of $\Delta$. Formally this is easily understood, as in
this case $\GG^\gtrless(t)=0$ and $\gamot(t)$ loses its dependence
on $g$ and $\Delta$, and thus the asymmetry is lost in the phonon
induced dephasing. One can interpret this approximation as only
allowing the phonons to interact with the bare electron and not the
electron-photon quasi-particle, the polariton, that is actually
present in the system. This illustrates the importance of
accounting for the polaritonic quasi-particle nature of the strongly
coupled QD-cavity and its non-Markovian interaction with the phonon
reservoir.

\begin{figure}[t]
  \includegraphics[width=0.4\textwidth]{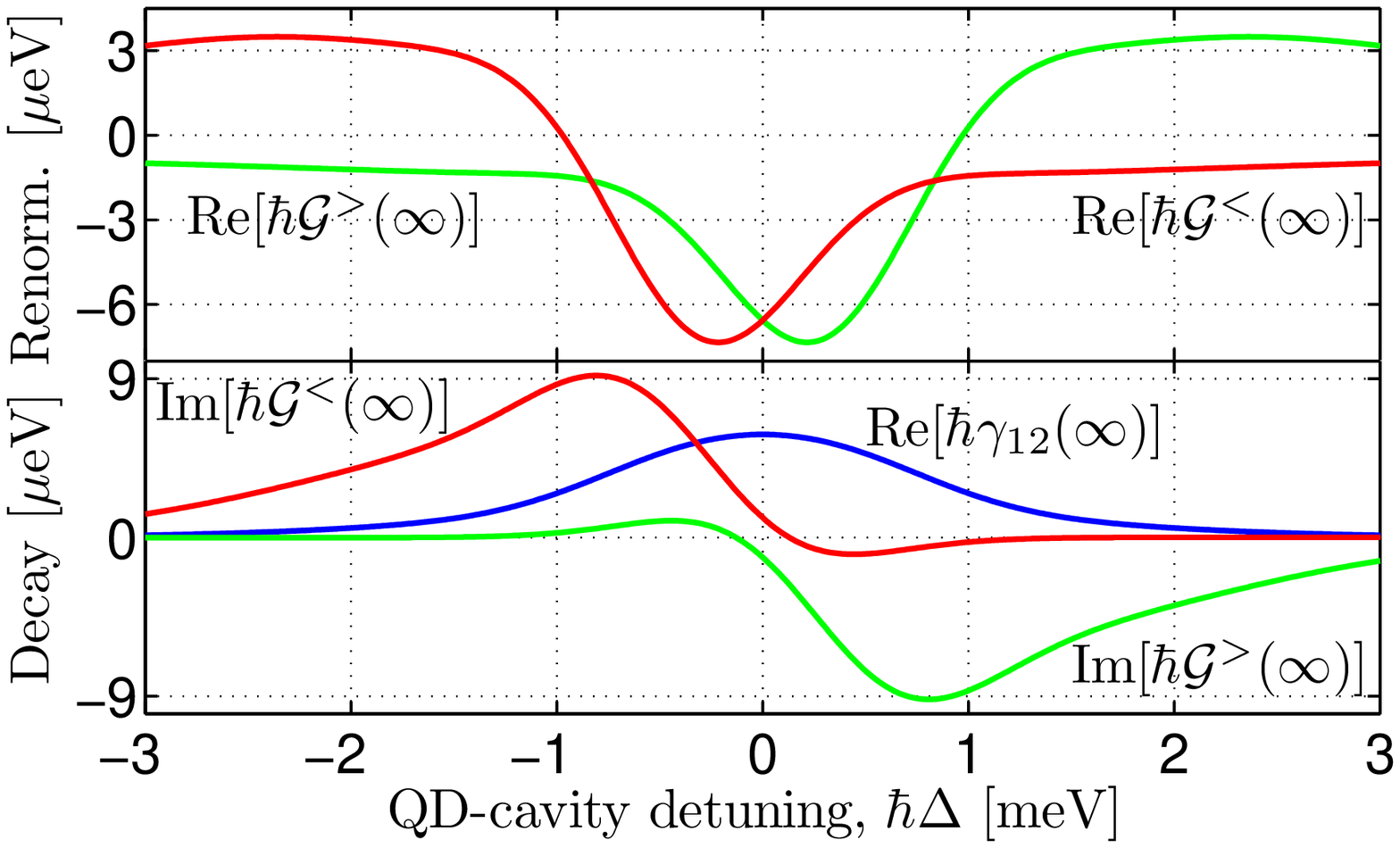}
\caption{(color online). $t \rightarrow \infty$ limit of $\hb\gamot(t)$ (blue) and $\hb\GG^\gtrless(t)$ (green/red) for the parameters $\hb g = 140~\mu$eV and $T = 4$ K. $\mrm{Im}[\hb\gamot(\infty)]$ has the constant value 32.7 $\mu$eV.}
  \label{fig:Markov_rates}
\end{figure}
Figure \ref{fig:Markov_rates} shows the long-time limit
\cite{MarkovNote} of the rates, 
Eqs. (\ref{eq:Ggl_def}) and (\ref{eq:gam12_def}). A strongly
asymmetric $\GG^\gtrless(\infty)$ is apparent versus detuning. Also,
$\GG^>(\infty)$ and $\GG^<(\infty)$ attain very different values for
fixed detuning. As the pure dephasing rate $\gamma_{12}(\infty)$
remains symmetric with respect to detuning, the asymmetry in Fig.
\ref{fig:Purcell_decay_curves} is caused by the renormalization
rates $\GG^\gtrless(\infty)$. It is remarkable that despite of their
relative weakness, $|\GG^\gtrless(\infty)|_\mrm{max}/g < 7~\%$, such
large dynamical effects may occur. The physical origin of this
asymmetry can be traced back to the phonon correlation functions
given in Eq. (\ref{eq:Dgtrless_def}). Here it is seen that phonon
absorption processes are suppressed at low temperatures as they are
proportional to the occupation factor $n_{\bs k}$, whereas phonon
emission processes continue to be possible due to the presence of
the phonon vacuum field. In linear QD absorption spectra
virtual phonon emission results in asymmetric
sidebands centered around an infinitely sharp zero phonon line (ZPL)
\cite{Krummheuer.PRB.2002.phonon_study}. However, for
the
present physical system it is essential that the
cavity field is treated non-linearly. This results in novel effects
such as a finite width of the ZPL and the coupling strength
renormalization rates $\GG^\gtrless(t)$.

\begin{figure}[t]
    \includegraphics[width=0.4\textwidth]{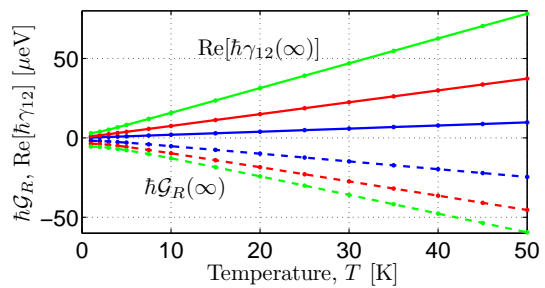}
\caption{(color online). Real part of $\hb\gamot(\infty)$ (solid) and $\hb\GG_R(\infty)$ (dashed) for $\hbar g = 50~\mrm{(blue)}, 100~\mrm{(red)},150~\mrm{(green)}$ $\mu$eV. The imaginary parts are independent of temperature and much smaller than the real part and therefore not shown.}
  \label{fig:Markov_rates_Temperature}
\end{figure}
We next examine temperature effects at zero detuning. For this case: $\GG^\gtrless(t,\Delta =0)=\GG_R(t)\mp i \GG_I(t)$,
where $\GG_R(t)$ and $\GG_I(t)$ are real functions. Figure \ref{fig:Markov_rates_Temperature} shows the effect of temperature on the phonon induced rates $\GG^\gtrless(\infty)$ and $\gamot(\infty)$, within the low temperature regime typically explored in cQED experiments. As expected, the pure dephasing rate increases with temperature. We also observe an increase as a function of the bare coupling strength $g$. $\GG_R(\infty)$ also increases in magnitude with increasing temperature, but has a negative value. This leads to a lowering of the effective coupling strength entering the equation for $s_{12}(t)$ as temperature is increased. This can be realized by inserting Eqs. (\ref{eq:s12_ph}) and $\GG^\gtrless(\infty,\Delta =0)$ into Eq. (\ref{eq:s12_eom}) yielding $g_\mrm{eff}(\infty)=g-|\GG_R(\infty)|<g$.
\begin{figure}[t]
  \includegraphics[width=0.4\textwidth]{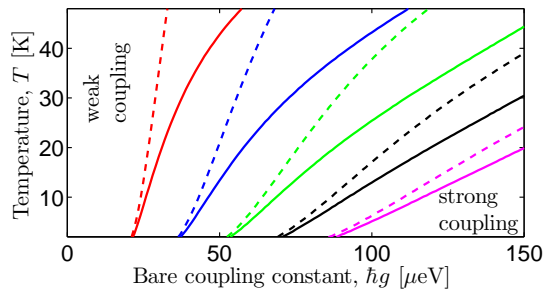}
\caption{(color online). Parameter space showing the presence of strong or weak coupling for the full model (solid) and for $\GG^\gtrless(t)=0$ (dashed), with $\hbar \kappa  =$ 75 (red), 125 (blue), 175 (green), 225 (black), 275 (magenta) $\mu$eV.}
  \label{fig:SCWC_crosss}
\end{figure}
The effective coupling strength thus depends significantly on
temperature. This mechanism is also relevant for interpreting cQED
absorption spectra \cite{Milde.PRB.2008}.
We expect this dependence to have a detrimental effect on the
possibility of reaching the strong coupling regime. To quantify this
prediction we have investigated
the transition between the weak and strong coupling regime, while
varying the most important parameters in the model, namely $g$,
$\kappa$, and $T$. We define the weak coupling regime as the
situation where the initially populated excited state of the QD
decays monotonically toward zero. Figure \ref{fig:SCWC_crosss} shows
the results both for the full model and for comparison the case
where we have artificially put $\GG^\gtrless(t)=0$. This is done to
emphasize the effect of the temperature induced renormalization of
$g$, by allowing only the pure dephasing rate to be temperature
dependent, as is common practice in phenomenological cQED models. As
expected, we generally observe the presence of strong coupling in
the system for large $g$ and low $T$, with the parameter space of
strong coupling becoming extended as we increase the quality of the
cavity. Comparing the results of the full model with those where
$\GG^\gtrless(t)=0$ we notice a strong effect of the renormalization
of the bare coupling constant $g$. The parameter space where strong
coupling is obtained is significantly decreased when including the
renormalization of $g$. This result is relevant in the
interpretation of experimental data, as state-of-the-art cQED models
\cite{PureDephPhen} do not include the renormalization effects
contained in the functions $\GG^\gtrless(t)$.
These effects are of significant importance and therefore cQED
models neglecting them can lead to misinterpretation of experimental
data.

In conclusion, we have illustrated the importance of applying a dressed-state picture of the polaritonic QD-cavity system when modeling cQED systems interacting with LA phonons. We have shown its relation to recent experiments, explaining the observed asymmetry in lifetimes with respect to the QD-cavity detuning. The asymmetry can only be understood by treating the QD-cavity system as a polaritonic quasi-particle in the phonon induced scattering terms. Furthermore we have investigated a phonon induced lowering of the effective coupling strength with increasing temperature, which was found to change the criterion for strong coupling significantly.

While completing this manuscript we became aware of a recent preprint with related work \cite{Ota.arXiv.2009}, however with focus on modeling experimental emission spectra as opposed to interpretation of the mechanism responsible for the dynamics, which is presented here.


\bibliographystyle{plain}

\begin{thebibliography}{}
\bibitem{phononMEdef}
The phonon matrix element is defined as $M^{\bs k}_\mrm{\nu\nu} = (2 d c_\mrm s V)^{-1/2}(\hbar k)^{1/2}D_\nu \int d\bs r |\phi_\nu(\bs r)|^2 \e{-i\bs k \cd \bs r}$, with $\phi_\nu(\bs r) \propto \exp[-(x^2+y^2)/(2l^2_{xy,\nu})] \cos (\pi z/l_\mrm{eff,z})$. We take typical GaAs parameters: $d=5370~\mrm{kg}\mrm{m}^{-3}$, $c_\mrm s=5110~\mrm{m}\mrm{s}^{-1}$, $D_\mrm e = -14.6~\mrm{eV}$, $D_\mrm g = -4.8~\mrm{eV}$, $l_{xy,\mrm{e}}=6.18~\mrm{nm}$, $l_{xy,\mrm{g}}=4.48~\mrm{nm}$, and $l_\mrm{eff,z}=8~\mrm{nm}$. $l_\mrm{eff,z}$ is to be considered as an effective length.
\bibitem{ExciteCon}
We note that our model describes both purely resonant and quasi-resonant excitation. In the latter case large intraband relaxation rates are assumed, while for smaller rates our main conclusions still hold.
\bibitem{MarkovNote}
The $t\rightarrow \infty$ limit is effectively reached after 5 ps for most parameters. The long time limit is not identical to the Markov limit, as this is only obtained when the reservoir has no memory.
\end{thebibliography}

\end{document}